\documentclass[]{sigir_forum}

\def\pubissue{Vol. 55 No. 1 - June 2021}

\def\pubtype{OPINION PAPER}

\begin{document}

\title{Rethinking Search: \\ Making Domain Experts out of Dilettantes\footnote{Disclaimer: This is a research proposal, not the roadmap for any Google product or service.}}

\author{
  Donald Metzler\\
  Google Research\\
  \emph{metzler@google.com}
\and
  Yi Tay\\
  Google Research\\
  \emph{yitay@google.com}
\and
  Dara Bahri\\
  Google Research\\
  \emph{dbahri@google.com}
\and
  Marc Najork\\
  Google Research\\
  \emph{najork@google.com}
}

\maketitle


\begin{abstract}
    When experiencing an information need, users want to engage with a domain expert, but often turn to an information retrieval system, such as a search engine, instead. Classical information retrieval systems do not answer information needs directly, but instead provide references to (hopefully authoritative) answers. Successful question answering systems offer a limited corpus created on-demand by human experts, which is neither timely nor scalable. Pre-trained language models, by contrast, are capable of directly generating prose that may be responsive to an information need, but at present they are dilettantes rather than domain experts -- they do not have a true understanding of the world, they are prone to hallucinating, and crucially they are incapable of justifying their utterances by referring to supporting documents in the corpus they were trained over. This paper examines how ideas from classical information retrieval and pre-trained language models can be synthesized and evolved into systems that truly deliver on the promise of domain expert advice.
\end{abstract}


\section{Introduction}
Given an information need, users often turn to search engines for help. Such systems point them in the direction of one or more relevant items from a corpus. This is appropriate for navigational and transactional intents (e.g. home page finding or online shopping) but typically less ideal for informational needs, where users seek answers to questions they may have~\citep{broderSF02}. Classical information retrieval (IR) systems do not directly answer information needs, but instead provide references to (hopefully authoritative) content.

The very fact that ranking is a critical component of this paradigm is a symptom of the retrieval system providing users a selection of potential answers, which induces a rather significant cognitive burden on the user. The desire to return \emph{answers} instead of ranked lists of results was one of the motivating factors for developing question answering systems. While there has been a great deal of research into QA systems, large-scale practical success has been somewhat limited. The original vision of question answering was to provide human-quality responses (i.e., ask a question using natural language and get an answer in natural language). Question answering systems have only delivered on the question part. Their responses, however, are either traditional lists of relevant documents, snippets extracted from documents, or answers created by human editors (e.g., Yahoo! Answers, Naver, Quora). While these solutions go beyond the experience afforded by classical IR systems, they suffer from a number of issues, including those related to coverage (e.g., answers are only provided for a small fraction of all possible questions) and authoritativeness (e.g., answers are often crowdsourced from both high and low quality sources). 

When it comes to natural language understanding, there has been significant progress over the past decade that has largely been fueled by the deep learning movement. Early advances, which have had wide-ranging impact across many research disciplines, include word embeddings (which capture word similarity) \citep{pennington2014glove,mikolov2013distributed}, advances in sequence modeling (which capture morphological and grammatical phenomena), and pre-trained language models (LMs) \citep{brownNIPS2020language,menaICLR18learning,radford2019language} (which can capture information about the relationship between entities). Improvements in these technologies are driven by ever-expanding data sets and model sizes, which allows such models to encode more and more knowledge about the world and demonstrate impressive capabilities to generalize via zero- and few-shot learning.

Unlike traditional IR or question answering systems, state-of-the-art pre-trained LMs \citep{devlin2018bert,brownNIPS2020language,raffel2020exploring} are capable of directly generating prose that may be responsive to an information need. However, such models are \emph{dilettantes} -- they do not have a true understanding of the world, they are prone to hallucinating, and crucially they are incapable of justifying their utterances by referring to supporting documents in the corpus they were trained over. This paper argues that many of these limitations result from the fact that such models fail to bridge the gap between sequences of terms and documents (and all the important meta-information associated with documents like provenance, authorship, authoritativeness, polarity, etc.).

Given the significant recent progress developing information retrieval, question answering, and pre-trained language modeling capabilities, now is an opportune time to take a step back to try to envision what possibilities the future might hold in terms of synthesizing and evolving these technologies into the next generation of IR systems that can help us get one step closer to truly domain expert quality responses.

This paper envisions how \emph{domain experts} can be created by leveraging pre-trained LMs. Of course, actual domain experts have a ``true understanding" of a given topic. Building such domain experts would likely require developing an artificial general intelligence, which is beyond the scope of this paper. Instead, by ``domain expert" we specifically mean that the system is capable of producing results (with or without actual ``understanding") that are of the same quality as a human expert in the given domain. To achieve this, this paper explores how ideas from classical IR and pre-trained LMs can be synthesized and evolved into systems that deliver on the promise of domain expert response quality. 

To move well beyond the current state-of-the-art, the fundamental assumptions that underlie modern IR systems need to be questioned. One of the key assumptions that this paper takes a critical look at is whether search indexes as we know them today are absolutely necessary or do they perhaps impose unnecessary and artificial restrictions on systems.

The inverted index has served as the workhorse of most modern search engines over the past several decades~\citep{croftBOOK09}. Such indexes encode term frequencies, term positions, document structure information, various forms of document metadata (e.g., document length), etc. They allow users to query the system using a mix of terms, phrases, and other so-called “advanced search operators” (e.g., “title:”). On the other hand, inverted indexes treat words as uninterpreted tokens, they do not capture their semantics. Specifically, the index is oblivious of morphology (instead, current IR systems perform stemming or lemmatization prior to indexing or retrieval), term similarity (instead, queries are expanded with synonyms prior to retrieval), or grammar (the closest thing to a LM captured by the index is word frequency distributions). 

Over the past few years, advances in representation learning resulted in a shift away from traditional inverted indexes towards dense vector-based indexes (or hybrid inverted + vector-based indexes)~\citep{gaoECIR21,karpukhin2020dense,khattabSIGIR20,kuziARXIV20,leeACL19,linARXIV20,xiongICLR2021approximate}. These indexes encode semantically-rich document representations that primarily help improve recall by overcoming the vocabulary mismatch problem that is known to plague inverted indexing-based systems.

Many language understanding advances have already been successfully leveraged by IR researchers. For example, representation learning has been used for retrieval purposes, pre-trained LMs are being leveraged for scoring, etc. These efforts have yielded significant improvements across a range of tasks.

Despite all of this progress, today's cutting edge IR systems are not fundamentally different than classical IR systems developed many decades ago. Indeed, a majority of today's systems boil down to: (a) building an efficient queryable index for each document in the corpus, (b) retrieving a set of candidates for a given query, and (c) computing a relevance score for each candidate. This \emph{index-retrieve-then-rank} blueprint has withstood the test of time and has rarely been challenged or seriously rethought. 

This paper envisions a consolidated model-based approach to building IR systems that eliminates the need for indexes as we know them today by encoding all of the knowledge for a given corpus in a model that can be used for a wide range of tasks. As the remainder of this paper shows, once everything is viewed through a model-centric lens instead of an index-centric one, many new and interesting opportunities emerge to significantly advance IR systems. If successful, IR models that synthesize elements of classical IR systems and modern large-scale NLP models have the potential to yield a transformational shift in thinking and a significant leap in capabilities across a wide range of IR tasks, such as document retrieval, question answering, summarization, classification, recommendation, etc.

\section{Related Work}
This section provides a brief survey of research related to document retrieval, question answering, knowledge bases, and pre-trained LMs, as those are the research directions that are most relevant and closely aligned to the envisioned system.

\subsection{Document Retrieval}
Document retrieval has a rich history. Rather than undertake a comprehensive literature review here, we instead focus on three specific lines of important recent research that have culminated in the current state-of-the-art document retrieval systems~\cite{mitraFNTIR18}.

The first such line of research is \emph{learning to rank}, which was propelled by the commercial success of search engines and easy access to large volumes of user interaction data. This movement represented a transformational leap beyond traditional TF.IDF-based IR systems. There is a vast and continually growing body of literature focused on this topic. Interested readers are encouraged to see \citep{liBOOK14} and \citep{liuFNTIR09} for more details.

The next line of research is \emph{neural-based re-ranking models}. This line of research can be thought of as a specific application of neural networks to the problem of learning to rank. These models typically take documents retrieved in some way (e.g., from a traditional inverted index or dense vector index) and use neural network-based models to score or rank documents. Some examples of such models include Deep Relevance Matching Model (DRMM)~\citep{guoCIKM16}, DUET~\citep{mitraWWW17}, Kernel-based Neural Ranking Model (KNRM)~\citep{xiongSIGIR17}, Position-Aware Convolutional-Recurrent Relevance (PACRR)~\citep{huiACL17}, and Context-Aware PACRR (co-PACRR)~\citep{huiWSDM18}. This is a highly active area of research, with continuous progress  as a result of newer, better modeling architectures, novel uses of data, etc. For more information on this topic, interested readers should see \citep{mitraFNTIR18} and \citep{onalIRJ17}.

The third and final line of research is \emph{representation learning}. The goal of representation learning is to encode queries and/or documents into (often dense) vector representations. These representations can be used for a variety of purposes including retrieval, for example via efficient $k$-nearest neighbor search. There have been many such approaches proposed in the literature~\citep{gaoECIR21,karpukhin2020dense,khattabSIGIR20,kuziARXIV20,leeACL19,linARXIV20,xiongICLR2021approximate}. One of the key benefits of these approaches over term-based representations is that the encodings often capture rich semantics and provide a way of overcoming the well-known vocabulary mismatch problem. However, one of the shortcomings of these approaches is that the recall improvements they bring often come at the cost of reduced precision compared to term-based representations.

The culmination of these three lines of research represent the current state-of-the-art retrieval systems~\cite{mitraFNTIR18}. These systems often rely on a combination of term-based (i.e., retrieval over an inverted index) and semantic (i.e., retrieval over an index of dense vector representations) retrieval to generate an initial set of candidates. This set of candidates is then typically passed into one or more stages of re-ranking models, which are quite likely to be neural network-based learning-to-rank models. As mentioned previously, the index-retrieve-then-rank paradigm has withstood the test of time and it is no surprise that advanced machine learning and NLP-based approaches are an integral part of the indexing, retrieval, and ranking components of modern day systems.

\subsection{Question Answering}
Early research into question answering systems primarily focused on ranking and retrieval, whereby models are trained to learn a relevance score between a user question and candidate answers \citep{wang2007jeopardy,yang2015wikiqa,severyn2015learning,tan2015lstm}. Due to the nature of how the task is defined, systems typically rely on advances in short text matching \citep{rao2019bridging}, paraphrase identification \citep{he2015multi}, and entailment detection \citep{parikh2016decomposable,tayEMNLP18compare}. More recently, neural network-based models have been designed for question answer matching and have made significant progress on the problem \citep{severyn2015learning,wangICLR2017compare,tay2018multi}. 

As time goes on, there has been a slight shift in how the question answering problem is expressed. The development of new neural network-based modules, pointer networks \citep{vinyals2015pointer}, and consequently the Match-LSTM with answer pointers \citep{wangICLR2017machine} have unlocked the potential for highly effective extractive question answering. Instead of ranking question-answer pairs, the new pointer mechanism enables the extraction of answer spans within passages. As such, this spurred significant growth in the number of models proposed for QA tasks of this sort~\citep{rajpurkar2016squad,trischler2017newsqa}. Likewise, a surge of neural network-based models, primarily attention-based \citep{yu2018qanet,wang2017gated,tay2018densely}, have been developed for tackling this question answering task variant.

The typical setup of machine reading comprehension (MRC) systems involve a query and a context (passage). In practice, these passages do not appear out of thin air, i.e., they have to be retrieved from somewhere. This motivated another variant of the QA problem which is commonly referred to as open domain question answering~\citep{joshi2017triviaqa,dhingra2017quasar,dunn2017searchqa}. Here, the goal is to first retrieve the relevant passages such that an MRC model can extract the correct answer~\citep{clarkACL2018simple}. To this end, there have been multiple innovations on this front, such as jointly learning or modeling interactions between the retrieval system and the MRC model~\citep{wang2018r,dasICLR2019multi}. Hence, retrieval still remains a core component of QA systems, especially when the corpus is large \citep{karpukhin2020dense}.

The final class of QA systems are generative ones. In retrieval and/or span-based QA systems, it is always assumed some notion of ground truth exists in either the provided passages or amongst the answer candidates. Generative QA systems shift this burden to the generative model whereby the only assumption is that the answer exists in the generator's output vocabulary. Historically, this has been thought of as significantly more challenging than extractive or retrieval-based QA~\citep{kovcisky2018narrativeqa,tanAAAI2018snet,tay2019simple}. Today, pre-trained encoder-decoder (seq2seq) models such as T5~\citep{raffel2020exploring} and BART~\citep{lewisACL20bart} have demonstrated that state-of-the-art QA performance can be achieved via generative models. 

\subsection{Explicit Knowledge Bases}
\label{sect:kb}

During the early 2000s, research momentum around the Semantic Web~\citep{bernersleeSCIAM01} and pre-existing ``old-style'' AI research gave rise to graph-structured knowledge bases, including Freebase~\citep{bollackerSIGMOD08}, WikiData~\citep{vrandevicCACM14}, the Google Knowledge Graph~\citep{singhalURL12}, and Microsoft Satori~\citep{qianURL13}. A knowledge base is typically realized as a set of triplets -- a pair of \emph{entities} and a \emph{predicate} relating them. The triplets induce a graph structure, with entities as the nodes and predicates as labeled edges. Knowledge graphs are well-suited to represent factoids (e.g. ``Thomas Edison invented the light bulb''), and query algebras over the graph structure make it possible to form short chains of relations. Originally assembled and curated by hand, there has been much research on automatically extracting knowledge graph triplets from Web pages, including Yago~\citep{suchanekWWW07}, NELL~\citep{carlsonWSDM10,mitchellCACM18}, and Knowledge Vault~\citep{dongKDD14}. Google leverages its Knowledge Graph when generating ``Knowledge Panels'' (cards containing a collection of factoids directly embedded in the results page) in response to a factoid-seeking query. These direct answers bring us some of the way towards our vision of domain expert advice; however, they are limited by the size of the graph, which only represents a fraction of the information contained in the Web corpus, and the inability to provide nuanced answers (by definition, answers are limited to factoids).

\subsection{Pre-Trained Language Models}
Over the past few years, pre-trained LMs have had a significant impact on the field of NLP. Models such as like BERT \citep{devlin2018bert}, RoBERTa \citep{liu2019roberta}, XLNet, T5 \citep{raffel2020exploring}, BART \citep{lewisACL20bart} GPT-2 \citep{radford2019language}, GPT-3 \citep{brownNIPS2020language}, and Meena \citep{adiwardana2020towards} are state-of-the-art for most (if not all) NLP tasks. The key idea behind pre-trained LMs is to first pre-train using one or more generative tasks, after which one may simply apply the pre-trained model to downstream tasks by fine-tuning their parameters. To this end, language modeling \citep{brownNIPS2020language}, masked language modeling \citep{devlin2018bert}, and encoder-decoder based generation \citep{raffel2020exploring} have proven to be highly effective pre-training approaches. 

One of the core reasons why pre-trained LMs are so successful is that they learn highly effective contextual representations. Research on learning contextual representations dates back to early work of learning semantic word vectors whereby models like SkipGram \citep{mikolov2013efficient,mikolov2013distributed} and GloVE \citep{pennington2014glove} helped revolutionize the field of NLP. Subsequently, pre-trained models such as CoVe \citep{mccann2017learned} and ELMo \citep{peters2018deep} have also demonstrated the benefits of more sophisticated pre-training objectives and model architectures. 

Today, pre-trained LMs are generally based on Transformer models \citep{vaswani2017attention}. Unlike predecessors that are trained largely on recurrent neural network models \citep{mccann2017learned,peters2018deep}, the Transformer's ability to be parallelized efficiently enables practitioners and researchers to greatly scale these models. Large-scale models have shown to generalize better, as shown in the zero- and few-shot experiments of \citep{brownNIPS2020language}, and achieve significantly better performance~\citep{raffel2020exploring}. Many of the largest models are billion-scale, with the largest T5 model reaching 11 billion parameters and GPT-3 reaching $175$ billion parameters. Very recently, Switch Transformers~\citep{fedus2021switch} broke through the \textit{trillion} parameter ceiling. 

Pre-trained LMs such as GPT-3 have demonstrated impressive text generation capabilities. In fact, some of the text synthesized by such models are indistinguishable from text written by humans~\citep{ippolito2019automatic}. 

\section{Model-Based Information Retrieval}
\label{sec:model}
We begin the more technical portion of the paper by posing the following questions:
\begin{itemize}
    \item What if we got rid of the notion of the index altogether and replaced it with a pre-trained model that efficiently and effectively encodes all of the information contained in the corpus?
    \item What if the distinction between retrieval and ranking went away and instead there was a single response generation phase?
\end{itemize}
Recent breakthroughs in natural language understanding (e.g., BERT), language modeling, few-shot learning, and multi-task learning (e.g., T5) provide supporting evidence that these questions are not as far-fetched as they may have been just a couple of years ago. Indeed, the confluence of these advances has created a unique opportunity to meaningfully explore answers to these questions.

\begin{figure}
    \centering
    \includegraphics[width=0.5\columnwidth]{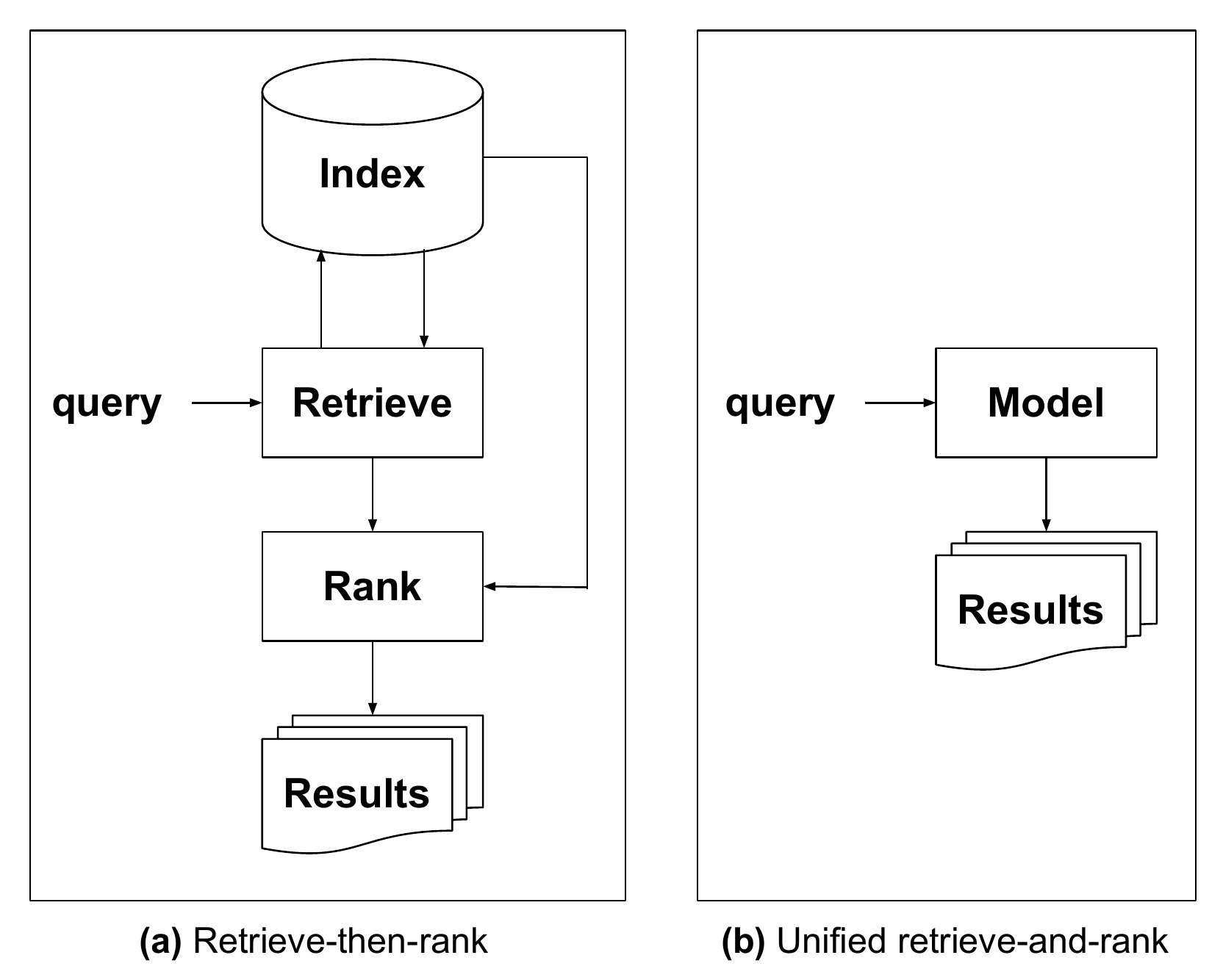}
    \caption{High-level schematics of the traditional index-retrieve-then-rank (left) and model-based (right) paradigms.}
    \label{fig:paradigms}
\end{figure}

This section describes a modeling approach that synthesizes aspects of modern IR systems and NLP models. The approach, referred to as \emph{model-based information retrieval}, is meant to replace the long-lived ``retrieve-then-rank" paradigm by collapsing the indexing, retrieval, and ranking components of traditional IR systems into a single consolidated model. With model-based IR, indexing is replaced with model training, while retrieval and ranking are replaced with model inference. See Figure~\ref{fig:paradigms} for a high-level schematic of these two paradigms. 

It is of course important to acknowledge that models are already used everywhere in modern IR systems. The important distinction between the systems of today and the envisioned system is the fact that a consolidated model replaces the indexing, retrieval, and ranking components. In essence, it is referred to as model-based because there is nothing but a model.

This represents a fundamentally different way of thinking about IR systems. Within the index-retrieve-then-rank paradigm, modeling work (e.g., query understanding, document understanding, retrieval, ranking, etc.) is done on top of the index itself. This results in modern IR systems being comprised of a disparate mix of heterogeneous models (e.g., one model used to learn document representations, another for document understanding, and yet another for ranking). Within the model-based information retrieval paradigm, the model and the index are one. Everything that was previously developed on top of the index is now integrated directly into a single, consolidated model. The model itself is built from the corpus, just like indexes are built from the corpus, but the encoded information is expected to be much more complex and able to solve a wider range of tasks. 

For example, for question answering tasks our envisioned model is able to synthesize a single answer that incorporates information from many documents in the corpus, and it will be able to support assertions in the answer by referencing supporting evidence in the corpus, much like a properly crafted Wikipedia entry supports each assertion of fact with a link to a primary source. This is just one of many novel tasks that this type of model has the potential to enable.

The following sub-sections dive deeper into some of the fundamental building blocks that are necessary for this model-based approach to be possible.

\subsection{Beyond Language Models}
Pre-trained LMs have proven to be useful for a wide range of NLP and IR tasks. However, such models fundamentally work on the term-level. Common natural language tasks, like masked language modeling, typically take a sequence of terms as input and produce one or more terms as output.

As the literature clearly demonstrates, there are many ways to represent queries and documents with such models. However, nearly all previously proposed work first tokenizes queries and/or documents into sequences of terms that are then passed as input to some model. The output of the model can then be used in a variety of ways. For example, embeddings can be used as learned representations, generated terms can be used to augment an inverted index, the models can be fine-tuned and used for ranking, and so on.

This approach is obviously quite useful, but it does have a number of limitations. LMs that are purely learned over sequences of terms have no way of explicitly modeling relationships between terms and documents. LMs essentially learn \emph{assertions} (``The sky is blue.") from the corpus they are trained over but fail to learn \emph{associations} between terms and individual documents. This is why we refer to pre-trained LMs as dilettantes -- they are perceived to know a lot but their knowledge is skin deep. 

Given that such models only know about sequences of terms, it is not possible to provide higher-level entities (like document ids) as input or expect document ids to be produced as output without making some changes to the underlying model. To replace indexes with a single, consolidated model, it must be possible for the model itself to have knowledge about the universe of document identifiers, in the same way that traditional indexes do. One way to accomplish this is to move away from traditional LMs and towards \emph{corpus models} that jointly model term-term, term-document, and document-document relationships.

Of course, modern LMs are trained over a corpus of word sequences, and therefore can be considered rudimentary corpus models. However, since such models are agnostic to higher-level corpus structure and document properties, they fall far short of being faithful models of a corpus.

Corpus models, as defined, can take as input a sequence of terms and output one or more terms or one or more document ids. Similarly, the model could take as input a mix of terms and document ids and output one or more terms and/or document ids. By explicitly encoding the associations between terms and documents, the model suddenly becomes able to ``natively" retrieve (and score) documents without the need for a traditional index.

How to actually build corpus models that are both efficient (at training and inference time) and effective is an open research question that spans multiple research communities. There are many obvious things that can be tried, such as adding a sentinel token to the vocabulary for each document identifier, but then the question immediately becomes how can one meaningfully pre-train such a model? Another option is to connect document identifiers to input or output sequences of terms using a separate model or procedure, which might be more scalable but is less consolidated as it would likely need to be done ``outside" of the consolidated model itself. An option is to investigate early work in learning document representations, i.e., doc2vec or paragraph2vec~\citep{mikolov2013distributed} that learns embeddings for documents by infusing document identifiers in the pre-training stage. However, this raises additional questions of how to incrementally update the index. Should additional training stages be incorporated so that a model may learn new document-term associations? 

Another research challenge is how to effectively scale the number of document identifier tokens. Document identifiers have to be allocated as extra ids in the output layers of the language model, which can substantially increase the number of model parameters. Clearly, when there is no upper bound on the total number of documents, as is often the case in dynamic corpora, this quickly becomes a concern. Some options include representing document identifiers as sequences of subwords (or characters), factorizing the id space in some way, or storing identifiers in some form of structured memory module.

Overall, this is an important and potentially highly impactful, but long-overlooked, line of research that could benefit both the IR and NLP communities.

\subsection{Multi-Task Learning: Towards A Single Model for all Information Retrieval Tasks}
We envision using the same corpus model as a multi-task learner for multiple IR tasks. To this end, once a corpus model has been trained, it can of course be used for the most classical of all IR tasks -- document retrieval. However, by leveraging recent advances in multi-task learning, such a model can very likely be applied to a diverse range of tasks.

By leveraging a multi-task text-to-text corpus model with appropriately defined pre-training objectives and fine-tuning tasks, one can envision a consolidated model approach to IR that can be used for document retrieval, question answering, summarization, and new tasks such as the aspirational task of providing domain expert advice that was described in the introduction.

The T5 model~\citep{raffel2020exploring} and follow-ups demonstrated that it is possible to achieve state-of-the-art performance across multiple tasks with a single consolidated model. The key idea is to leverage task conditioning via a task identifier that tells the model which task it is supposed to perform. The T5 model has been shown to achieve state-of-the-art on several challenging language understanding benchmarks. Hence, it is expected that a sufficiently high quality corpus-based model trained in a similar manner would be capable of equally strong performance across multiple tasks of interest.

\begin{figure}
    \centering
    \includegraphics[width=0.7\columnwidth]{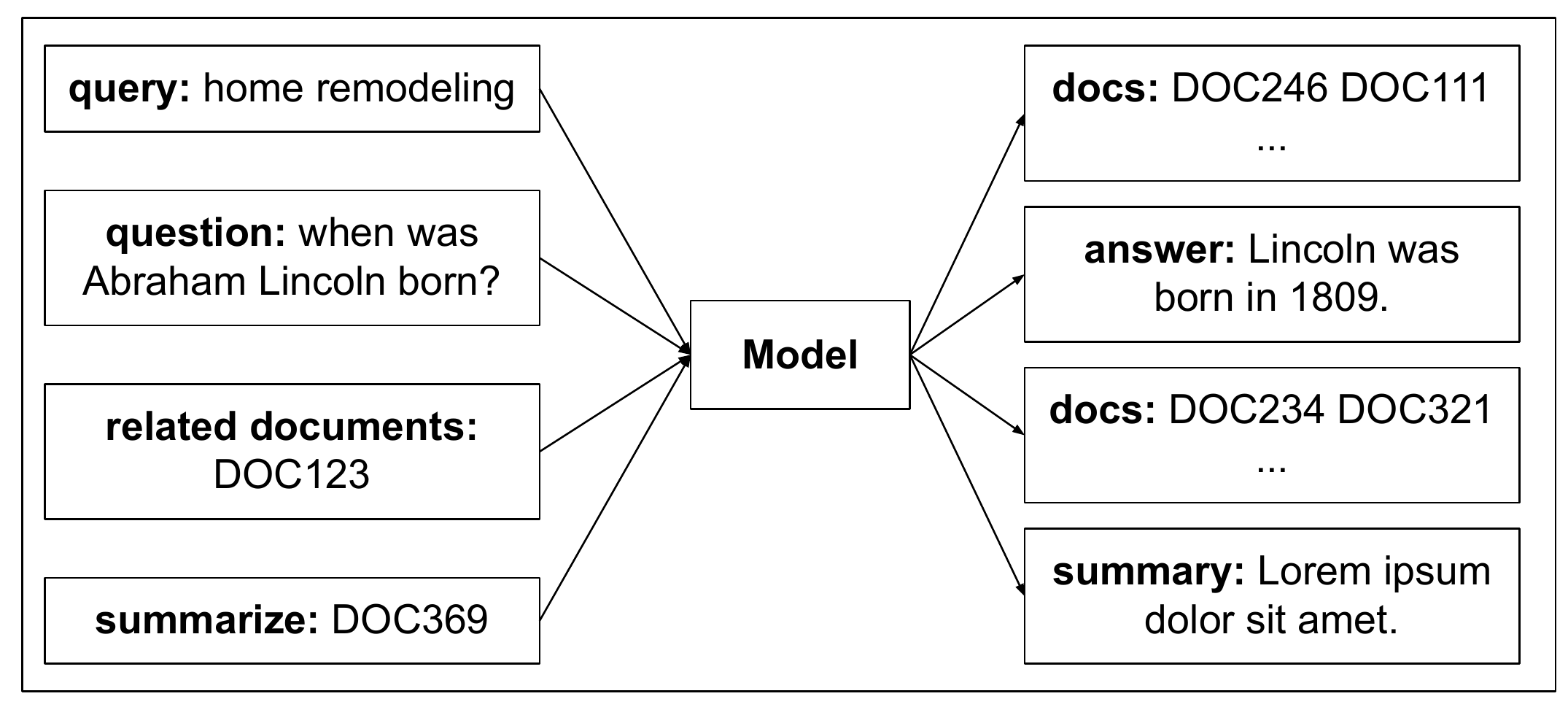}
    \caption{Example of how a single consolidated model can be leveraged to solve a wide range of IR tasks. This example shows a model that handles document retrieval, question answering, related document retrieval, and document summarization tasks.}
    \label{fig:unified-model}
\end{figure}

Figure~\ref{fig:unified-model} demonstrates what this might look like from a practical perspective where the input to such a consolidated model is a task-prefixed request and the output is a response that satisfies the request. In this figure, note that the model is able to perform tasks defined over mixtures of terms and document identifiers. Such a setup would provide significantly more flexibility than the pure term-based LMs that are widely used today.

The tasks in this figure include:
\begin{itemize}
    \item \textbf{Document retrieval.} The input is a query string and the output is one or more relevant document identifiers.
    \item \textbf{Question answering.} The input is a question in natural language and the output is a natural language response.
    \item \textbf{Related document retrieval.} The input is a document identifier and the output is a one or more relevant document identifiers.
    \item \textbf{Document summarization.} The input is a document identifier and the output is a summary of the document.
\end{itemize}
These are obviously only for illustrative purposes and there is no limit to the types of tasks that could potentially be incorporated into such a model.

Moving towards a consolidated model for all of IR opens up a number of potentially interesting and impactful research directions that span machine learning, NLP, and IR.

\subsection{Zero- and Few-Shot Learning}
Another advantage of pre-trained models is their ability to perform well in zero- and few-shot learning situations. This is particularly appealing for tasks where limited training data is available. Indeed, many real-world IR systems have access to very little in the way of labeled data. For this reason, being able to generalize well based on a small number of labeled examples has the potential to yield significant practical utility.

Zero- and few-shot learning is common in document ranking tasks. For example, \emph{ad hoc} retrieval can be thought of as a zero-shot learning task since no examples of relevant documents for the given query are actually provided to the system. Furthermore, relevance feedback can be thought of as few-shot learning, since the user manually provides labels for one or more documents that the system can use to improve its ranking.

Building upon the general consolidated modeling paradigm developed in the previous sub-sections, these tasks can easily be defined as follows:

\paragraph{Ad Hoc Retrieval (zero-shot)}
\begin{itemize}
    \item \textbf{Input:} $query$
    \item \textbf{Output:} $reldoc_1, \ldots, reldoc_n$
\end{itemize}
where $query$ is a query string and $reldoc_i$ are document identifiers.

\paragraph{Pseudo-relevance feedback (few-shot)}
\begin{itemize}
    \item \textbf{Input:} $(query_1, doc_1), \ldots, (query_n, doc_n)$ $query$
    \item \textbf{Output:} $reldoc_1, \ldots, reldoc_n$
\end{itemize}
where $(query_i, doc_i)$ are pairs of query strings and document identifiers that have been labeled as relevant in some way and $reldoc_i$ are document identifiers. In this way, the labeled query/document pairs are provided as context to the system to enable few-shot learning for the current $query$.

Beyond document retrieval, consolidated models can be used in a few-shot learning setting for other tasks, including query understanding and document understanding. For example,

\paragraph{Query Understanding (few-shot)}
\begin{itemize}
    \item \textbf{Input:} $(query_1, intent_1), \ldots, (query_n, intent_n)$ $query$
    \item \textbf{Output:} $intent$
\end{itemize}
where $(query_i, intent_i)$ are pairs of query strings and intents (categories) that have been labeled in some way. These are passed as context to the model, which then uses them to generalize to identify the best $intent$ associated with $query$.

\paragraph{Document Understanding (few-shot)}
\begin{itemize}
    \item \textbf{Input:} $(doc_1, label_1), \ldots, (doc_n, label_n)$ $doc$
    \item \textbf{Output:} $label$
\end{itemize}
where $(doc_i, label_i)$ are pairs of document identifiers and document labels. The model then takes $doc$ as input and generates $label$ as output.

As these examples show, having a consolidated multi-task model that understands the connections between sequences of terms and document identifiers opens up a wide range of straightforward and powerful use cases, even when there is only limited labeled data available, in an extremely straightforward manner.

\subsection{Response Generation}
Using a T5-like setup or more generally any encoder-decoder model, it is possible to leverage the model to generate a wide range of potential output types. As described in the previous sub-sections, these outputs could be sequences of terms, document identifiers learned as part of a consolidated corpus model, query intent or document category labels learned as a result of fine-tuning or few-shot learning, and so on.

\begin{figure*}
    \centering
    \includegraphics[width=\textwidth]{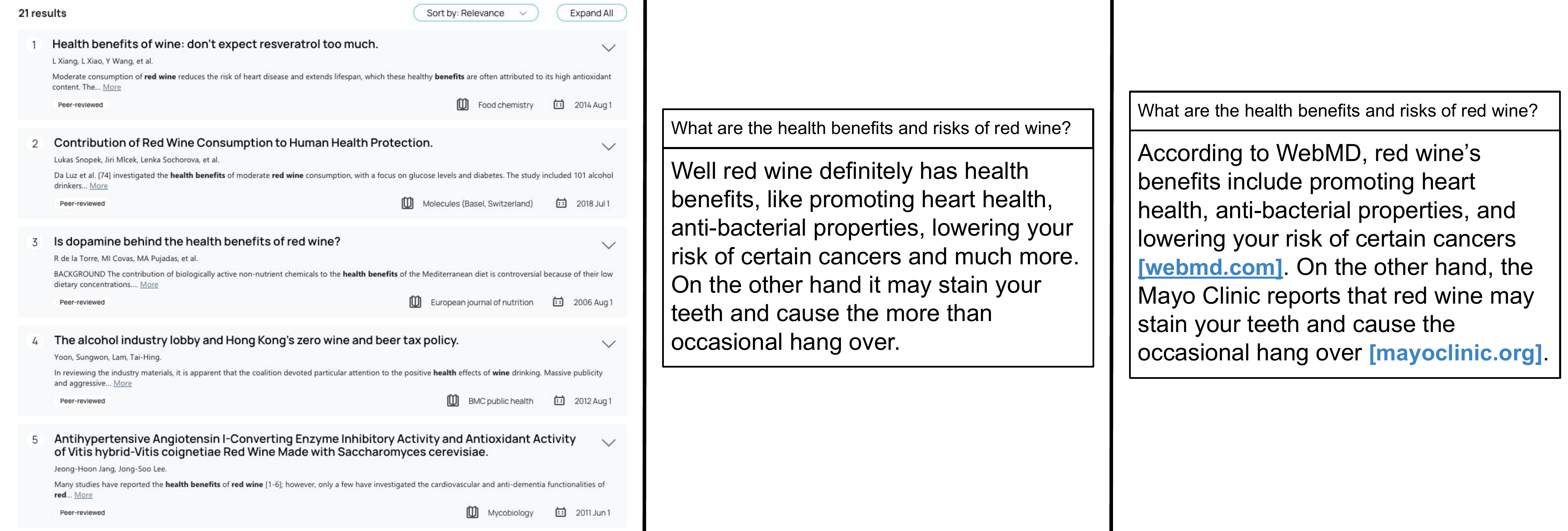}
    \caption{Example domain-specific search engine (left), pre-trained language model (middle), and envisioned domain expert (right) responses for the query ``What are the health benefits and risks of red wine?".}
    \label{fig:examples}
\end{figure*}

An aspirational goal would be a retrieval system that, given the query ``what are the health benefits and risks of red wine," would give you a coherent and authoritative answer laying out the evidence for both benefits and risks. Figure~\ref{fig:examples} shows the responses returned by a domain-specific search engine (left) and a modern pre-trained LM for this example query. The search engine returns a number of relevant documents and provides query-biased snippets. On the other hand, the pre-trained LM returns a coherent, focused response that seemingly answers the question but provides no context for the answer or any sense of how authoritative or comprehensive it is. The system envisioned in this paper would instead be able to produce a response like the one on the right. This response provides references to source material making it much easier to highlight the authoritativeness of the content. This simple example shows how deeper connections between sequences of words and documents can be useful.

There are many possible ways to approach this problem. The model itself can understand both terms and documents and their relationships and be trained to generate content with proper citations. This alone is a major challenge in terms of how to define the task, where labeled (or weakly labeled) data might be sourced from, how to evaluate the output, etc. Another possibility is to use a standard pre-trained LM and add a \emph{learning to cite} task on top of it that can be used to artificially ground synthesized text to articles in the corpus. This solution has a number of drawbacks, including the fact that the generation and citation processes are disjoint and hence may lead to incongruous outputs. On the other hand, jointly performing generation and citation will likely be more challenging but is likely to yield better results. A potential approach could perhaps be in similar spirit to learning a mixture of distributions \citep{see2017get,mccann2018natural} and adaptively learning to toggle between generating document identifiers and raw tokens.

There are also a number of other major challenges associated with generating high quality responses. A response is high quality if it exhibits the following properties:
\begin{itemize}
    \item \textbf{Authoritative.} Responses should generate content by pulling from highly authoritative sources. This is another reason why establishing more explicit connections between sequences of terms and document metadata is so crucial. If all of the documents in a corpus are annotated with an authoritativeness score, that score should be taken into account when training the model, generating responses, or both.
    \item \textbf{Transparent.} Whenever possible, the provenance of the information being presented to the user should be made available to them. Is this the primary source of information? If not, what is the primary source?
    \item \textbf{Unbiased.} Pre-trained LMs are trained to maximize their predictive power on their training data, and thus they may reflect societal biases in that data~\citep{benderFAT21,hutchinsonACL20,shengEMNLP19}. To address those risks, designers of systems that employ pre-trained LMs may consider different training objectives~\citep{WebsterARXIV20} and also surround the model with additional safeguards against biased system responses. 
    \item \textbf{Diverse perspectives.} Generated responses should represent a range of diverse perspectives but should not be polarizing. For example, for queries about controversial topics, both sides of the topic should be covered in a fair and balanced way. This obviously has close tie-ins with model bias.
    \item \textbf{Accessible.} Written in terms that are understandable to the user. For example, responses that discuss complex medical issues should be written in as-plain-as-possible terms. Another example is authoritative content that may only be written in a certain language that is different than the one that the user issued their query in. In this situation, the system should provide a translated version of the response to the user.
\end{itemize}
This list is obviously not exhaustive but hopefully drives home the point that extreme care must to be taken to ensure that synthesized responses are indeed high quality. Doing so will require a significant amount of research across multiple disciplines. Even simply defining a measure of synthesized answer quality that takes into account all of these factors (and more) is itself an important but difficult research challenge. Building these principles into the model will be even more challenging.

\subsection{Reasoning Capabilities}
One key advantage of adopting a model-based index is that we can leverage the modularity of neural networks for composing new modules with specialized forms of inductive bias. While most pre-trained LMs today are based on the Transformer model architecture, such models may be augmented by composing them with one or more additional neural modules. For example, we may imbue the model with reasoning-like capabilities by allowing the model to attend over an external memory. To this end, neural modules that provide a memory-like inductive bias to enable memory lookups \citep{westonICLR2015memory,miller2016key} or content addressing (e.g., differentiable neural computers \citep{graves2016hybrid}) may be explored. Other forms of inductive biases such as multi-hop reasoning~\citep{chen2019multi,asaiICLR2020learning,zhaoICLR2020transformer} might also be useful.

Within the context of neural retrieval from an encoder-decoder model, it may also be possible to incorporate relational inductive biases \citep{asaiICLR2020learning,deNAACL19question} to model relationships amongst candidate documents and terms in the decoder. For instance, when learning to output document identifiers, the decoder learns what is already being partially generated and performs self-attention across the partial generation. While a simple way of thinking of this is through the lens of listwise learning-to-rank, there is significantly more flexibility in incorporating relational reasoning as a neural component where the model learns this in a data-driven manner. Conversely, it is not possible to develop systems that exhibit these reasoning-like properties with traditional indexes and classical IR models.

While the ability to reason is a nice characteristic for such models to have, it may also result in unfavorable outcomes. Specifically, sequence-to-sequence neural network-based models are prone to hallucination~\citep{leeIRASL18}. Hallucination has the potential to generate novel truthful outputs, but it also has the potential for to result in strange, untrue, or downright offensive outputs as well. This is an issue that is prevalent across all modern pre-trained LMs and one that will need to be addressed in some way (e.g., via the system outputting logical explanations) before their outputs can be trusted in a similar way as a human domain expert.

\subsection{Arithmetic, Logical, Temporal, and Geographical Reasoning}
It is well established that modern search engines can handle queries that deal with some form of arithmetic reasoning. For example, converting across currencies, i.e., \textit{`36,500 USD to pounds\'}, \textit{2am PST to GMT-2'} and \textit{`how far is California from New York City`}. Current search engines behave as if they have some sense of order, time, logic, and even geographical distance. While there has been recent work that investigates these aspects in the context of neural network models~\citep{ran2019numnet}, it remains challenging to develop neural models that can reliably and accurately deliver on these types of reasoning capabilities. Moreover, at a foundational level, LMs that can handle numerical or temporal reasoning can be quite crucial even for document understanding, as some form of numerical commonsense may be required to fundamentally understand the content of documents.

\subsection{Combining Modalities in a Single Model}
Another key advantage of the model-based paradigm is that it allows multiple modalities to be combined within a single model. Documents traditionally contain a significant amount of metadata and/or media content, such as images, videos, and audio. Traditionally, image search and document search leverage very different indexes. Having a consolidated model capable of handling multiple modalities can bridge this gap.

There has also been progress in vision-based Transformers~\citep{dosovitskiy2020image} and vision-based T5 \citep{cho2021unifying}. Such models provide a means for exploring multi-modal grounding as a way of enabling effective text and image representations. This typically involves having a separate encoder for each modality, making it straightforward to integrate into existing models.

Other modalities of interest include tabular data and traditional features (e.g., document metadata) that are passed to standard machine learning models. These supplementary features may be generated from another network (e.g., embeddings) or handcrafted. Here, it is an open research question how to integrate these auxiliary features into these pre-trained models.



\subsection{Leveraging Document and Corpus Structure}
Successful modern IR systems fully leverage all of the rich structure associated with documents. For example, terms that appear in the title or anchor text are often treated as more important for Web search applications. Today's modern pre-trained LMs fail to take this rich document structure into account. How to successfully model and leverage rich document structure is an interesting direction of future research that could provide significant benefit to IR-centric applications of pre-trained LMs.

In an open corpus such as the web, not all documents are equally authoritative or trustworthy. There are many known techniques for estimating the authority or veracity of a Web page, from fact-checking claims within a single page~\citep{jiangWWW2020} to aggregating quality signals at the logical domain level~\citep{dongPVLDB2015}. Incorporating such documents authority signals, along with other signals such as polarity or toxicity~\citep{wulczynWWW2019} of the content, into a language model is crucial to ameliorating biases that such models are prone to learn from unvetted documents ``in the wild''. 

Furthermore, many corpora have some form of explicit or implicit graph structure associated with them~\citep{broder2000graph}. Modern IR systems leverage such graphs in a number of ways, such as computing graph-based measures like PageRank~\citep{brinWWW98}, identifying hubs and authorities~\citep{kleinbergJACM99}, and so on. There are many ways that this structure can also be leveraged within the proposed framework. For example, the graph structure can be used for cross-document pre-training (i.e., finding similar documents and packing them into the same sequence for pre-training~\citep{caciularu2021cross}), thereby enabling the model to benefit from long-range dependencies and cross-document language understanding. Another potential way to leverage the graph structure is to define a co-training task that predicts if there is an edge between two documents. How to best leverage corpus structure within pre-trained LMs is an important and interesting open research question.

\subsection{Scaling to Multiple Languages}
Another key research question is whether it is possible to model all documents across languages within a single model. Practically, this has implications both in terms of vocabulary and model capacity. If this can be effectively addressed, the envisioned model would be able to support cross-lingual generalization and be applied to tasks like cross-language IR~\citep{nieBOOK10}. Early work has shown that this is already possible, given the success of multilingual T5~\citep{xue2020mt5} and other multilingual pre-trained models~\citep{pires2019multilingual}. Despite these early success, many important research challenges remain, such as determining the optimal proportions of training data from each language to use to effectively learn balanced representations.

\subsection{On the Challenges of Scale}
The modeling efforts outlined in this paper can be generally regarded as incredibly resource intensive. There are multiple perspectives to the challenge of scale of this overall endeavour. 

Firstly, there is question of model capacity and exactly how large of a model would be required to fit multiple tasks, billions of documents, document identifiers across a dozen of languages. We postulate that models need to go beyond a billion parameters to be effective in having enough capacity. However, models with large parameter footprints are difficult to serve in practice. 

Secondly, documents are generally long, easily spanning a thousand or more subwords. Additionally, modeling multiple documents can incur additional substantial costs. The problem of scale within the context of long document understanding with pre-trained LMs is a well-established problem \citep{beltagy2020longformer,tay2020efficient}. Dozens of models have been proposed to solve this issue but still sacrifice performance for memory efficiency~\citep{tay2020long}.

We believe that there are multiple research directions that enable us to scale. In terms of modeling capacity, a potential solution is to leverage models with a large number of parameters (e.g., trillions of parameters) but maintain the computation cost and inference time of a model an order of magnitude smaller. One good example of a recent model along these lines is the Switch Transformer \citep{fedus2021switch}. Models that leverage dynamic and conditional computation to select and activate certain sub-networks may be key to allow such systems to scale. These models also fit into the overall paradigm of modeling multiple tasks in a single network - since intuitively a model should only select the relevant sub-network for certain tasks. 

Complementing the challenges around scaling model size are those around \emph{sample efficiency}. For a model to be truly knowledgeable, it must be trained over a diverse distribution of data. However, training on large and diverse data sets may be infeasible in practice. Techniques that can ``condense'' or ``distill'' massive training datasets into smaller, more manageable ones, with little loss in information will likely need to be employed~\citep{wang2018dataset,zhaoICLR2021dataset}.

\subsection{Incremental Learning}
There are many research and engineering challenges associated with keeping such models up-to-date in the presence of potentially highly dynamic corpora. For example, it is an open question as to how models can be built in a manner such that it is efficient (and effective) to add new documents to the model. ``Online'' or ``incremental'' learning explores the problem of updating machine learned models as new data arrives sequentially in a way that does not harm performance on previous data, a phenomenon known ``catastrophic forgetting'' \citep{french1999catastrophic}. The ``continual learning'' setup generalizes this and studies models and techniques for learning on new tasks without forgetting old ones. While many methods have been proposed (see \citep{parisi2019continual,deTPAMI21continual} for a survey), it has mostly been studied on toy datasets and synthetic setups in a low-parameter count regime. Investigating whether current methods work on pre-trained language models remains an open and important research direction.

Even more interestingly and more challenging is the problem of having models ``forget" everything that they know about a document that was removed from the corpus. This becomes even more challenging in situations where privacy or legal reasons require that all traces of a deleted piece of content be removed from a system, which is a typical requirement when building practical IR systems.

\subsection{Model Interpretability, Controllability, and Robustness}
Since the operating mechanism of classical term-based IR systems is transparent to designers, how the system will behave on test queries is often predictable. Deviations from desired behavior are easier to debug and can even be fixed by manually adding new rules, although such manual interventions and hard-coded rules are hard to scale. In contrast, it is well-known that modern deep neural networks suffer from interpretability issues, and addressing them is an active line of research (e.g. \citep{sundararajan2017axiomatic}; see \citep{guidotti2018survey} for a survey). Furthermore, even after an issue with the model has been identified, it is often unclear what modeling knobs one should turn to fix the model's behavior. A desiderata then is that the model should be both interpretable and debuggable as well as \emph{controllable}, i.e. the model designer should know how to control the behavior of the trained model, e.g. by modifying training data or tuning hyper-parameters in the loss function. Of equal importance is robustness. For example, should the search user make the benign typo: ``the'' $\rightarrow$ ``teh'' in an otherwise good query, we expect that the model's response will not change drastically. Crucially, we would like the model to be well-behaved for queries it may not have seen before, including adversarial examples \citep{goodfellowICLR15explaining} that can occur not due to malicious intent by the user but rather by bad luck.

\subsection{Putting It All Together}
If all of these research ambitions were to come to fruition, the resulting system would be a very early version of the system that we envisioned in the introduction. That is, the resulting system would be able to provide domain expert answers to a wide range of information needs in a way that neither modern IR systems, question answering systems, or pre-trained LMs can do today.

Some of the key benefits of the model-based IR paradigm described herein include:
\begin{itemize}
    \item It abstracts away the long-lived, and possibly unnecessary, distinction between “retrieval” and “scoring”.
    \item It results in a consolidated model that encodes all of the knowledge contained in a corpus, eliminating the need for traditional indexes.
    \item It allows for dozens of new tasks to easily be handled by the model, either via multi-task learning or via few-shot learning, with minimal amounts of labeled training data.
    \item It allows seamless integration of multiple modalities and languages within a consolidated model.
\end{itemize}

\section{Conclusions}
This paper envisions an ambitious research direction that doubles down on the synthesis between modern IR and NLP to deliver on the long-promised goal of providing human expert quality answers to information needs. Specifically, the paper makes the case for developing retrieval systems that combine the best elements of document retrieval systems and pre-trained language models. To accomplish this, a so-called model-based information retrieval framework is proposed that breaks away from the traditional index-retrieve-then-rank paradigm by encoding the knowledge contained in a corpus in a consolidated model that replaces the indexing, retrieval, and ranking components of traditional systems. It was argued that if successful, such a consolidated model can be used to solve a wide range of tasks (via multi-task learning), can easily adapt to new low resource tasks and corpora (via zero- and few-shot learning), and can be used to synthesize high quality responses that go well beyond what today's search and question answering systems are capable of.

There are a number of interesting and difficult research and engineering challenges that must be solved before the envisioned system can be realized. These challenges span the IR, NLP, and machine learning research disciplines, and will require interdisciplinary research to be successful. Some of the major challenges include modeling (moving from LMs to corpus model), training (pre-training objectives, fine-tuning task definitions), response generation (authoritativeness, bias mitigation), and scalability (indexing and serving).

\bibliography{main.bib}

\end{document}